\documentclass[lettersize,journal]{IEEEtran}
\usepackage{amsmath,amsfonts}
\usepackage{algorithmic}
\usepackage{algorithm}
\usepackage{array}
\usepackage[caption=false,font=normalsize,labelfont=sf,textfont=sf]{subfig}
\usepackage{textcomp}
\usepackage{stfloats}
\usepackage{url}
\usepackage{verbatim}
\usepackage{graphicx}
\usepackage{tikz}
\usepackage{tikz/tikz_definitions}
\usepackage{circuitikz}

\usepackage{math_definitions}
\usepackage{tudacolors}
\usepackage{pgfplotstable}
\usepgfplotslibrary{colormaps} 
\usetikzlibrary{pgfplots.colormaps} 
\usetikzlibrary{decorations.markings}
\usetikzlibrary{patterns}
\makeatletter
\tikzset{
  nomorepostactions/.code={\let\tikz@postactions=\pgfutil@empty},
  mymark/.style 2 args={decoration={markings,
    mark= between positions 0 and 1 step (1/25)*\pgfdecoratedpathlength with{%
        \tikzset{#2,every mark}\tikz@options
        \pgfuseplotmark{#1}%
      },  
    },
    postaction={decorate},
    /pgfplots/legend image post style={
        mark=#1,mark options={#2},every path/.append style={nomorepostactions}
    },
  },
}
\makeatother
\usepackage{graphicx}
\usepackage{tikzHelper}
\usepackage{siunitx}
\usepackage{hhline}

\usepackage[backend=biber, style=ieee]{biblatex}

\AtEveryBibitem{%
  \ifentrytype{online}{%
  }{%
    \clearfield{howpublished}%
    \clearfield{url}%
    \clearfield{urldate}%
    \clearfield{issn}%
  }%
}

\makeatletter
    \tikzset{scale line widths/.style={%
    /utils/exec=\pgfgettransformentries{\tmpa}{\tmpb}{\tmpc}{\tmpd}{\tmp}{\tmp}%
    \pgfmathsetmacro{\myJacobian}{sqrt(abs(\tmpa*\tmpd-\tmpb*\tmpc))}%
    \pgfmathsetlength\pgflinewidth{\myJacobian*0.4pt}%
    \def\tikz@semiaddlinewidth##1{\pgfmathsetmacro{\my@lw}{\myJacobian*##1}%
    \tikz@addoption{\pgfsetlinewidth{\my@lw pt}}\pgfmathsetlength\pgflinewidth{\my@lw pt}},%
    thin}}
\makeatother

\begin{document}

\title{Magneto-Thermal Thin Shell Approximation for 3D Finite Element Analysis of No-Insulation Coils}

\author{Erik Schnaubelt, Sina Atalay, Mariusz Wozniak, Julien Dular, Christophe Geuzaine, Benoît Vanderheyden, Nicolas Marsic, Arjan Verweij, and Sebastian Schöps
\thanks{The work of Erik Schnaubelt has been sponsored by the Wolfgang Gentner Programme of the German Federal Ministry of Education and Research (grant no. 13E18CHA) and by the Graduate School Computational Engineering within the Centre for Computational Engineering at the Technical University of Darmstadt. (\emph{Corresponding author: E. Schnaubelt}.)}
\thanks{E. Schnaubelt is with CERN, 1211 Meyrin, Switzerland and the Technical University of Darmstadt, 64289 Darmstadt, Germany (e-mail: erik.schnaubelt@cern.ch).}
\thanks{S. Atalay is with CERN, 1211 Meyrin, Switzerland and Boğaziçi University, 34342 Bebek/Istanbul, Türkiye.}
\thanks{M. Wozniak, J. Dular and A. Verweij are with CERN, 1211 Meyrin, Switzerland.}
\thanks{S. Schöps and N. Marsic are with the Technical University of Darmstadt, 64289 Darmstadt, Germany.}
\thanks{C. Geuzaine and B. Vanderheyden are with the University of Liège, 4000 Liège, Belgium.} 
}

\markboth{}%
{}

\makeatletter
\def\ps@IEEEtitlepagestyle{
  \def\@oddfoot{\mycopyrightnotice}
  \def\@evenfoot{}
}
\def\mycopyrightnotice{
  {\footnotesize
  \begin{minipage}{\textwidth}
  This work has been submitted to the IEEE for possible publication. Copyright may be transferred without notice, after which this version may no longer be accessible.
  \end{minipage}
  }
}


\maketitle

\begin{abstract}
For finite element (FE) analysis of no-insulation (NI) high-temperature superconducting (HTS) pancake coils, the high aspect ratio of the turn-to-turn contact layer (T2TCL) leads to meshing difficulties which result in either poor quality mesh elements resulting in a decrease of the solution accuracy or a high number of degrees of freedom. We proposed to mitigate this issue by collapsing the T2TCL volume into a surface and using a so-called thin shell approximation (TSA). Previously, two TSA have been introduced, one to solve the heat equation and the other for an $\vec{H}-\phi$ magnetodynamic formulation.

In this work, we propose to combine the magnetodynamic and thermal TSA to create a coupled magneto-thermal TSA for three-dimensional FE analysis. Particular attention is paid to the detailed derivation of the coupling terms. 
In the context of NI HTS pancake coils, the TSA represents the electric and thermal contact resistance of the T2TCL. For the HTS coated conductor (CC) itself, an anisotropic homogenization is used which represents its multi-layered structure. In axial and azimuthal direction, it resolves the current sharing between the HTS and other layers of the CC. The coupled TSA formulation is verified against a reference model with volumetric T2TCL. The coupled TSA is shown to significantly reduce the solution time as well as the manual effort required for high-quality meshes of the T2TCL. The implementation is open-source and a reference implementation is made publicly available.
\end{abstract}


\section{Introduction}

No-insulation (NI) pancake coils \cite{Hahn2011} have no turn-to-turn electrical insulation and are popular due to their high thermal stability \cite{hahn2018} resulting from a possibility for currents to bypass local normal zones \cite{Wang2015, Hahn_2016}. Despite this, quenches are still possible in NI coils \cite{Kim_2017, YANAGISAWA201440}. To this end, quench detection and protection of NI coils require appropriate modeling and analysis methods.  

Most commonly, simulations of NI coils are based on (distributed) network models; see \cite{hahn2018} for a summary. For FE-based simulations, two-dimensional (2D) axisymmetric techniques based on a homogenization employing an anisotropic resistivity tensor have been proposed in 2020 \cite{Mataira_2020}. More recently, in 2023, 2D axisymmetric models based on the Minimum Electro Magnetic Entropy Production (MEMEP) method have been studied \cite{pardo2023fast}.

However, NI coils commonly exhibit true three-dimensional (3D) geometries and quenches are intrinsically local effects. To comprehensively simulate quench, full 3D models are thus desirable. These are difficult due to the current flow across the turn-to-turn contact layer (T2TCL) \cite{Mataira_2020}, which has a high aspect ratio \cite{Schnaubelt_2023aa}. In \cite{Wang_2021}, classical FE models with volumetric T2TCL have been used to study the AC loss of an NI pancake coil. They were, however, restricted to sinusoidal sources and lacking the thermal coupling. To ensure an accurate solution, the T2TCL requires a high number of degrees of freedom (DoF) in a classical FE method. 

Recently, we proposed to collapse the T2TCL volumes to surfaces using a so-called thin shell approximation (TSA). First, a thermal TSA to represent thin insulation layers by considering the heat equation was introduced in \cite{Schnaubelt_2023aa}. Second, a magnetodynamic $\vec{H}-\phi$ TSA was used to study the charge and discharge of NI coils in \cite{Schnaubelt_2023ab}. By replacing T2TCL volumes with surfaces, no volumetric mesh of the thin layer is required. This work proposes to combine the thermal TSA of \cite{Schnaubelt_2023aa} and the magnetodynamic TSA of \cite{Schnaubelt_2023ab} to model T2TCL with magneto-thermal T2TCL taking into account the coupling via \emph{i)} non-linear material relations and \emph{ii)} Joule losses.

One-dimensional (1D) Lagrange elements are used to discretize both temperature and magnetic field strength across the thickness of the T2TCL \cite{Alves_2021aa, Alves2022}. As discussed in \cite{dular2020}, the choice of formulation is important for robust and efficient simulations of systems with high-temperature superconductors (HTS). Since no ferromagnetic material exists in the computational domain, the $\vec{H}-\phi$ formulation is preferred over $\vec{A}$ formulations. Special care is needed to treat multiply connected domains, which is achieved using automatically created cohomology basis functions \cite{pellikka2013}. The HTS coated conductor (CC) is approximated by using a homogenization with anisotropic material properties which represent the multi-layered structure of the CC. In particular, the current sharing between the HTS and other layers is resolved.  

The magneto-thermal TSA is verified by comparison against a volumetric T2TCL model. The TSA is shown to be a robust alternative which produces accurate solutions with significantly reduced solution time and meshing effort. The source code with the details of the formulation is made available \cite{codeRepo}.

Section~\ref{sec:formulaiton} presents the volumetric T2TCL model with the classical FE formulation. It is replaced with surface T2TCL with the coupled magneto-thermal TSA formulation in Section~\ref{sec:tsacoupled} with the derivation of the coupling terms.
The model parameters of a powering cycle simulation of an NI coil with local defect and implementation details are summarized in Section~\ref{sec:setup}. The results of these simulations are detailed in Section \ref{sec:numExpCoupled}. The major findings are summarized in Section \ref{sec:conclusion}.

\section{Magneto-Thermal Formulation}
\label{sec:formulaiton}
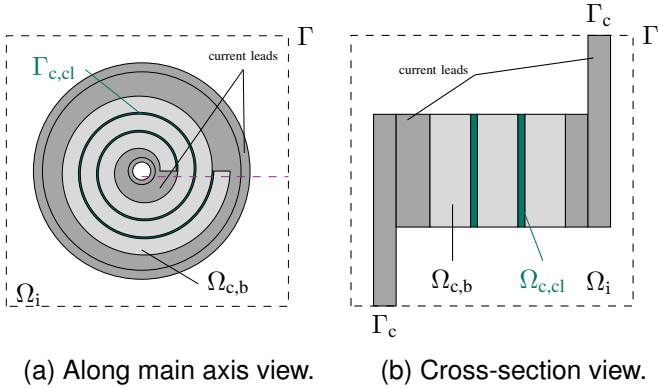
\begin{figure}[tbh]
     \subfloat[Along main axis view.]{
        \centering
        \scalebox{1}{\begin{tikzpicture}[scale=0.3, scale line widths]

    \begin{scope}[scale=0.8, scale line widths]
        \path [draw=none,fill=gray!70, even odd rule] (0,0) circle (2) (0,0) circle (0.5);
        \draw[black] (0, 0) circle(0.5);
        \draw[black] (0, 0) circle(0.75);
    
        \path [draw=none,fill=gray!70, even odd rule] (0,0) circle (6) (0,0) circle (3.5);
        \draw[black] (0, 0) circle(5.5);
        \draw[black] (0, 0) circle(6);
    
        \spiral[gray!30, line width=0.9cm](0,0)(0:720)(1.45:4.45)[1];
    
        \spiral[TUDa-3d, line width=0.1cm](0,0)(0:360)(1.95:3.95)[1];
    
        \spiral[black](0,0)(0:720)(1:4)[1];
        \spiral[black](0,0)(0:720)(1.9:4.9)[1];
    
        \draw[black] (1, 0) -- (2, 0); 
        \draw[black] (3.9, 0) -- (4.9, 0);
   \end{scope}


    \draw[dashed] (-6, -6) rectangle (6.5, 6) node[anchor=west]{$\Gamma$}; 

    \draw[TUDa-10b, dashed, very thick] (0, -0.25)  -- (6.5, -0.25);


    \node[] at (-5, -5.5) {$\Omega_\mathrm{i}$}; 
    
    
    \draw[black] (0, -3.5) -- (2.5, -5) node[right, black]{$\Omega_\text{c,b}$}; 
    \draw[TUDa-3d] (0, 2.5) -- (-2.5, 4.5) node[left, TUDa-3d]{$\Gamma_\text{c,cl}$}; 
    
    \draw[] (0.8, -0.766666666665) -- (4.5, 4.5) node[above]{\tiny current leads}; 
    \draw[] (4.5465291045174, 0.8016757497027) -- (4.5, 4.5) ; 

    \node[below] at (-4.5, -6) {\phantom{$\Gamma_\text{c}$}};
    \node[above] at (4, 6) {\phantom{$\Gamma_\text{c}$}};
\end{tikzpicture}}
      }%
    \subfloat[Cross-section view.]{
        \centering
        \scalebox{1}{\begin{tikzpicture}[scale=0.3]

    \fill[gray!70, draw=black] (-5, -6) rectangle (-4, 2.5); 
    \fill[gray!70, draw=black] (-4, -2.5) rectangle (-2.5, 2.5);

    \node[below] at (-4.5, -6) {$\Gamma_\text{c}$};
    \node[above] at (5, 6) {$\Gamma_\text{c}$};

   \fill[gray!70, draw=black] (3.5, -2.5) rectangle (4.5, 2.5); 
   \fill[gray!70, draw=black] (4.5, -2.5) rectangle (5.5, 6); 

   \fill[gray!30, draw=black] (-2.5, -2.5) rectangle (-0.7, 2.5); 
   \fill[gray!30, draw=black] (-0.4, -2.5) rectangle (1.4, 2.5); 
   \fill[gray!30, draw=black] (1.7, -2.5) rectangle (3.5, 2.5); 

   \fill[TUDa-3d, draw=black] (-0.7, -2.5) rectangle (-0.4, 2.5); 
       \fill[TUDa-3d, draw=black] (1.4, -2.5) rectangle (1.7, 2.5); 
   
    \draw[dashed] (-6, -6) rectangle (6.5, 6) node[anchor=west]{$\Gamma$}; 


    \draw[] (-0.5, 4.25) node[left, yshift=2, xshift=1]{\tiny current leads} -- (4.75, 5) ; 
    \draw[] (-0.5, 4.25) -- (-3.5, 2.25) ; 

    \node[] at (5, -5) {$\Omega_\mathrm{i}$};

    \draw[black] (-1.5, -1.5) -- (-1.5, -4) node[below, black]{$\Omega_\text{c,b}$}; 
    \draw[TUDa-3d] (1.55, -1.5) -- (2.5, -4) node[below, TUDa-3d]{$\Omega_\text{c,cl}$}; 
    
\end{tikzpicture}}
    }
    \caption{Computational domain $\Omega$ of the pancake coil with exterior boundary $\partial \Omega = \Gamma$. It consists of an insulating domain $\Omega_\text{i}$ and a conducting domain $\Omega_\text{c}$ which is divided into the bare CC $\Omega_\text{c,b}$, the T2TCL $\Omega_\text{c,cl}$ and current leads. The cross-section view in (b) is taken at the red dotted line shown in (a).}
    \label{fig:compDomain}
\end{figure}

The computational domain $\Omega$ as depicted in Fig.~\ref{fig:compDomain} consists of a conducting domain $\Omega_\text{c}$ and a non-conducting domain $\Omega_\text{i}$. The conducting domain consists of the bare homogenized CC $\Omega_\text{c,b}$, the T2TCL $\Omega_\text{c,cl}$ as well as the current leads. It is bounded by $\partial \Omega_\text{c}$ with outward normal vector $\vec{n}_\text{c}$.
The boundary of the domain is denoted by $\partial \Omega = \Gamma$ with outward normal vector $\vec{n}$. Furthermore, $\Gamma_\text{c}$ denotes the terminals, i.e., the surfaces where the current leads coincide with $\Gamma$, which allow a current or a voltage source to be imposed.

The weak formulation of the coupled magnetodynamic \cite{pellikka2013} and thermal \cite[Section 6.1.3.]{Ern2004} problem is: From a solution at time $t = 0$, find $\vec{H} \in \Hcurl$  and $T \in \Hgrad$ s.t.
\begin{align}
    \begin{split}
          \left(\kappa \gradient T, \gradient T'\right)_{\Omega_\text{c}} &+  \left( C_\text{V} \, \partial_t T, T' \right)_{\Omega_\text{c}} \\
         & = \left( \rho \, \vec{J} \cdot \vec{J}, T' \right)_{\Omega_\text{c}} \, \forall T' \in \Hgradzero,
    \end{split} \label{eq:thermal} \\
    \begin{split}
         \left(\partial_t \! \left(\mu \vec{H} \right), \vec{H}' \right)_{\Omega} &+ \left(\rho \curl \vec{H}, \curl \vec{H}' \right)_{\Omega_\text{c}} \\
         & = 0 \quad \forall \vec{H}' \in \Hcurlzero.  
     \end{split} \label{eq:mqs}
\end{align}
Herein, $\vec{H}$ is the magnetic field strength in \si{\ampere \per \meter}, $T$ the temperature in \si{\kelvin}, $\kappa$ the thermal conductivity in \si{\watt \per \meter \per \kelvin}, $C_\text{V}$ the volumetric heat capacity in \si{\joule \per \kelvin \per \cubic\meter}, $\rho$ the electric resistivity in \si{\ohm \meter}, $\vec{J} = \curl \vec{H}$ the electric current density in \si{\ampere \per \meter\squared} and $\mu$ the magnetic permeability in \si{\henry \per \meter}. The volume integral in $\Omega$ of the scalar product of the two arguments is denoted by $(\cdot,\cdot)_\Omega$. The coupling between \eqref{eq:thermal} and \eqref{eq:mqs} appears in the Joule loss term, i.e., the right hand side of \eqref{eq:thermal} as well as in the temperature and field dependencies of the materials. 

We considered the boundary conditions 
\begin{align}
    \vec{n} \times \vec{E} =  0 \quad \text{on} \quad \Gamma \quad \overset{\text{\cite{dular2000}}}{\Rightarrow}  \quad \vec{n} \cdot \left( \mu \vec{B} \right) = 0\quad \text{on} \quad \Gamma, 
    \label{eq:bc}
\end{align}
with the electric field $\vec{E}$ in \si{\volt \per \meter}, the magnetic flux density $\vec{B} = \mu \vec{H}$ in \si{T} and 
\begin{align}
    T &= g \quad \text{on} \quad \Gamma_\text{c}, \label{eq:dir}\\ 
    \vec{n}_\text{c} \cdot \left( \kappa \gradient T \right) &= 0 \quad \text{on} \quad \partial \Omega_\text{c} \setminus \Gamma_\text{c},
\end{align}
imposing the temperature on the terminals and adiabatic conditions everywhere else. Furthermore, $\Hcurl$ is the subspace of $H(\text{curl}, \Omega)$ with vanishing curl in $\Omega_\text{i}$ and strongly imposed source currents via cohomology basis functions \cite{pellikka2013}. Its subspace with zero current is $\Hcurlzero$. The subspace of $H^1(\Omega_\text{c})$ which fulfills the Dirichlet condition \eqref{eq:dir} is denoted as $\Hgrad$. Its subspace with $g = 0$ is $\Hgradzero$.

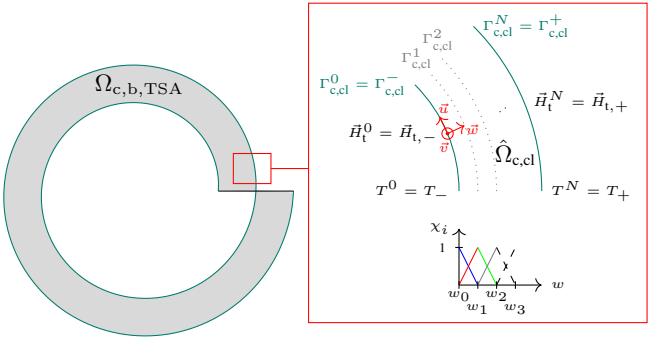
\begin{figure}[tbh]
    \centering
    \begin{tikzpicture}

    \begin{scope}[scale=1]
        \spiral[gray!30, line width=0.5cm](0,0)(0:0)(1.3:1.8)[1];

        \spiral[TUDa-3d](0,0)(0:360)(1.05:2.05)[1];
        \draw[] (1.05, 0) -- (1.55, 0); 
        \draw[] (1.55, 0) -- (2.05, 0); 
            
        \node[] at (0, 1.4) {\small $\Omega_\mathrm{c,b,TSA}$};

    \end{scope}
    
    \begin{scope}[shift={(4.25, 0)}]
        \draw[TUDa-3d] (0,0) node[left, align=center, black] {\tiny $T^0 = T_{-}$} arc (0:45:2) node[left, midway, align=center, black] {\tiny $\vec{H}_{\text{t}}^0 = \vec{H}_{\text{t},-}$} node[left, align=center] {\tiny $ \Gamma_\text{c,cl}^{0} = \Gamma_\text{c,cl}^{-}$}; 
        \draw[dotted, gray] (0.25,0) arc (0:45:2.25) node[anchor=south east, gray, xshift=5, yshift=-3] {\tiny $
        \Gamma_\text{c,cl}^{1}$}; 
        \draw[dotted, gray] (0.5,0) arc (0:45:2.5) node[anchor=south east, gray, xshift=8, yshift=-1] {\tiny $
        \Gamma_\text{c,cl}^{2}$}; 
        
        \draw[dotted] (0.5406687144, 1.052379439) -- (0.67596773944, 1.1480502971);  
        
        %
      \draw[->, red] (-0.15224093497, 0.76536686473) -- (  0.074336, 0.871021 ) node[midway,right] {\tiny $\vec{w}$}; 
       
      \draw[->, red] (-0.15224093497, 0.76536686473) -- ( -0.25790, 0.99194) node[midway, above, yshift=1] {\tiny $\vec{u}$}; 
       
      \fill[red] (-0.15224093497, 0.76536686473) node[anchor=north east, xshift=4] {\tiny $\vec{v}$} circle(0.025); 
       
        \draw[red] (-0.15224093497, 0.76536686473) circle(0.075); 
    
        \draw[TUDa-3d] (1.1, 0) node[right, align=center, black] {\tiny $T^N = T_{+}$}  arc (0:45:3.1) node[midway, right, align=center, black] {\tiny $\vec{H}_{\text{t}}^N = \vec{H}_{\text{t},+}$}  node[right, align=center] {\tiny $\Gamma_\text{c,cl}^{N} = \Gamma_\text{c,cl}^{+}$}; 
        
        \draw [draw=red] (-2,-1.75) rectangle (2.5,2.5);; 

            \node[] at (0.75, 0.5) {\small $\hat{\Omega}_\text{c,cl}$}; 

        \begin{scope}[shift={(0, -0.25)}]

            \draw[->] (0, -1) -- (1.1, -1) node[right] {\tiny $w$}; 
            
            \draw[->] (0, -1) -- (0, -0.25) node[left]{\tiny $\chi_i$};

            \draw[blue] (0, -0.5) -- (0.25, -1); 
            
            \draw[red] (0, -1) -- (0.25, -0.5); 
            
            \draw[green] (0.25, -0.5) -- (0.5, -1); 
            
            \draw[gray] (0.25, -1) -- (0.5, -0.5); 
    
            \draw[dashed] (0.5, -0.5) -- (0.75, -1); 
            
            \draw[dashed] (0.5, -1) -- (0.75, -0.5); 
            
            \draw[] (0, -1.05) node[below, yshift=2.5]{\tiny $w_0$} -- (0, -0.95); 
    
            \draw[] (0.25, -1.05) node[below, yshift=-2.5]{\tiny $w_1$} -- (0.25, -0.95); 
    
            \draw[] (0.5, -1.05) node[below, yshift=2.5]{\tiny $w_2$} -- (0.5, -0.95); 
    
            \draw[] (0.75, -1.05) node[below, yshift=-2.5]{\tiny $w_3$} -- (0.75, -0.95);

            \draw[] (-0.05, -0.5) node[left]{\tiny 1} -- (0.05, -0.5); 

        \end{scope}
    \end{scope}
    
    \draw [draw=red] (1.25,0.1) rectangle (1.75,0.5);
            
    \draw [red] (1.75,0.3) -- (2.25, 0.3);

\end{tikzpicture}
    \caption{{One turn of the HTS pancake coil (top view): for the TSA approach, the T2TCL is represented by a virtual domain $\hat{\Omega}_\text{c,cl}$ in which an internal FE discretization is used to solve the magneto-thermal problem.}}
    \label{fig:virtDisc}
\end{figure}

To represent the layered structure of the CC, an anisotropic resistivity is used inside $\Omega_\text{c,b}$. A local coordinate system $(\vec{u},\vec{v}, \vec{w})$ is introduced as shown in Fig.~\ref{fig:virtDisc} with $\vec{u}$ in tangential direction along the spiral winding, $\vec{v}$ in axial direction and $\vec{w}$ normal to the spiral winding. In the local coordinate system, the resistivity reads
\begin{equation}
    \rho_{uvw}|_{\Omega_\text{c,b}} = \text{diag} \left(\rho_{uu}, \rho_{vv}, \rho_{ww} \right).
\end{equation}
In $\vec{u}$- and $\vec{v}$-direction, the layers of the CC are electrically connected in parallel with the equivalent resistance
\begin{equation*}
    \rho_{uu} = \rho_{vv} = \left(\frac{f_\text{NC}}{\rho_{\text{NC}, uu}} + \frac{f_\text{HTS}}{\rho_{\text{HTS}}}\right)^{-1}, \label{eq:homRho} 
\end{equation*}
with $f_\text{HTS}$ the volume fraction of superconducting material in the CC, $f_\text{NC} = 1 - f_\text{HTS}$ the fraction of non-superconducting materials (here, copper, silver, and Hastelloy\textsuperscript{\textregistered}) and $\rho_{\text{NC}, uu}$ the equivalent parallel resistivity of non-superconducting materials. \textcolor{black}{The resistivity of the HTS is given by the power law, i.e., 
\begin{equation}
    \rho_{\text{HTS}} = \frac{E_\text{c}}
     {J_\text{c}}
     \left(  
      \frac{\lVert \vec{J}_\text{HTS} \rVert}{\, J_\text{c}} \right)^{n-1},
      \label{eq:powerLaw}
\end{equation}
with the critical electric field $E_\text{c}~=~\SI{E-4}{\volt \per \meter}$ and the fit of the critical current density $J_\text{c}\left( \lVert\vec{B} \rVert, T, \theta\right)$ from \cite{steam_material_library} with $\theta$ the angle between the tape wide surface and the $\vec{B}$-field. The current flowing in the HTS layer $\vec{J}_\text{HTS}$ reads
\begin{equation}
    \vec{J}_\text{HTS} = \frac{\lambda}{f_\text{HTS}} \left(\vec{J} \cdot \vec{u} \right) \vec{u} + \frac{\lambda}{f_\text{HTS}} \left( \vec{J} \cdot \vec{v} \right) \vec{v} + \left( \vec{J} \cdot \vec{w} \right) \vec{w},
\end{equation}
where the current sharing index $\lambda$ is the fraction of current flowing in the HTS layer. A non-linear root-finding problem needs to be solved as detailed in \cite[Section III.D]{Bortot_2020aa} and \cite{Schnaubelt_2023ab} to find $\lambda$. All other parameters in \eqref{eq:powerLaw} are presented in Table~\ref{tab:coilparams}.}

\textcolor{black}{In $\vec{w}$-direction, the equivalent series resistivity reads
\begin{equation*}
    \rho_{ww} = f_\text{NC} \, \rho_{\text{NC}, ww} + f_{\text{HTS}} \, \rho_{\text{HTS}}, \label{eq:homRhoAzimuthal} 
\end{equation*}
with $\rho_{\text{NC}, ww}$ the equivalent resistivity of the series connection of the non-superconducting materials.
In this work, $J_\text{c}$ and consequently $\rho_\text{HTS}$ are assumed to be the same in $\vec{w}$ and $\vec{u}-\vec{v}$ direction, but different functions could be used.}

The material tensor $\rho_{uvw}$ needs to be represented in the Cartesian $xyz$ coordinate system by using the transformation matrix from the local $uvw$ to the Cartesian coordinates 
\begin{equation}
    M = \left[\vec{u}(x, y, z), \vec{v}(x, y, z), \vec{w}(x,y,z) \right],
\end{equation}
with $\vec{u}$, $\vec{v}$ and $\vec{w}$ understood as row vectors. In Cartesian coordinates, the resistivity tensor then reads
\begin{equation}
    \rho|_{\Omega_\text{c,b}} = \rho_{xyz}|_{\Omega_\text{c,b}} = M \, \rho_{uvw}|_{\Omega_\text{c,b}} \, M^\top.
\end{equation}

The non-linear system is linearized using a quasi {Newton-Raphson} scheme where the derivatives of $\rho_\text{c,b}$ w.r.t. $\vec{J}$ is approximated using a finite difference scheme while all other derivatives are neglected.

The discretization of the magnetodynamics problem for pancake NI coils is discussed in detail in \cite{Schnaubelt_2023ab} while details for the thermal problem are found in \cite{Schnaubelt_2023aa}. The coupled problem is not discretized in a single monolithic linear system but two linear systems are created, one for \eqref{eq:thermal} and one for \eqref{eq:mqs}, which are then solved sequentially inside the Newton-Raphson loop.

\section{Thin Shell Formulation}
\label{sec:tsacoupled}
In order to treat the T2TCL in a magneto-thermal setting, we propose to couple the TSA proposed in \cite{Schnaubelt_2023aa} and \cite{Schnaubelt_2023ab}. In order to explain the coupling in detail afterwards, a summary of these previous papers is presented here. The main focus, however, is on the coupling of the two TSA.

\begin{figure}[tbh]
    \centering
    \begin{tikzpicture}[scale=0.3, scale line widths]
       \begin{scope}[scale=0.8, scale line widths]
        \path [draw=none,fill=gray!70, even odd rule] (0,0) circle (2) (0,0) circle (0.5);
        \draw[black] (0, 0) circle(0.5);
        \draw[black] (0, 0) circle(0.75);
    
        \path [draw=none,fill=gray!70, even odd rule] (0,0) circle (6) (0,0) circle (3.5);
        \draw[black] (0, 0) circle(5.5);
        \draw[black] (0, 0) circle(6);
    
        \spiral[gray!30, line width=0.9cm](0,0)(0:720)(1.45:4.45)[1];
    
        \spiral[TUDa-3d, line width=0.1cm](0,0)(0:360)(1.95:3.95)[1];
    
        \spiral[black](0,0)(0:720)(1:4)[1];
        \spiral[black](0,0)(0:720)(1.9:4.9)[1];
    
        \draw[black] (1, 0) -- (2, 0); 
        \draw[black] (3.9, 0) -- (4.9, 0);
   \end{scope}

    
    
    
    \draw[black] (0, -3.5) -- (2.5, -5) node[right, black]{$\Omega_\text{c,b}$}; 
    \draw[TUDa-3d] (0, 2.5) -- (-2.5, 4.5) node[left, TUDa-3d]{$\Omega_\text{c,cl}$};

  \draw[->] (5.75, 0) -- (7.75, 0) node[above, midway]{TSA}; 
    
    \begin{scope}[shift={(13, 0)}]

    \begin{scope}[scale=0.8, scale line widths]
        \path [draw=none,fill=gray!70, even odd rule] (0,0) circle (2) (0,0) circle (0.5);
        \draw[black] (0, 0) circle(0.5);
        \draw[black] (0, 0) circle(0.75);
    
        \path [draw=none,fill=gray!70, even odd rule] (0,0) circle (6) (0,0) circle (3.5);
        \draw[black] (0, 0) circle(5.5);
        \draw[black] (0, 0) circle(6);
    
        \spiral[gray!30, line width=1cm](0,0)(0:720)(1.5:4.5)[1];
    
        \spiral[TUDa-3d, thick](0,0)(0:360)(2:4)[1];
    
        \spiral[black](0,0)(0:360)(1:2)[0];
        \spiral[black](0,0)(0:360)(4:5)[0];

        \draw[black] (1, 0) -- (2, 0); 
        \draw[black] (4, 0) -- (5, 0);
   \end{scope}

    
    
    
    \draw[black] (0, -3.5) -- (2.5, -5) node[right, black]{$\Omega_\text{c,b}$}; 
    \draw[TUDa-3d] (0, 2.6) -- (-2.5, 4.5) node[left, TUDa-3d]{$\Gamma_\text{c,cl}$};

    \end{scope}
\end{tikzpicture}
    \caption{{HTS pancake coil (top view): for the TSA approach, the T2TCL volume $\Omega_\text{c,cl}$ (left) is collapsed into the T2TCL surface $\Gamma_\text{c,cl}$ (right).}}
    \label{fig:top_view_tsa}
\end{figure}
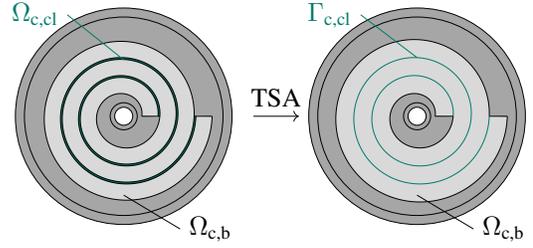

First, the volumetric T2TCL $\Omega_\text{c,cl}$ is replaced by a surface $\Gamma_\text{c,cl}$ as shown in Fig.~\ref{fig:top_view_tsa}. As $\Omega_\text{c,cl}$ is thermally and electrically conducting, the temperature and tangential magnetic field strength are discontinuous in order to represent temperature gradients and surface current densities. As proposed in \cite{Geuzaine_2000}, this discontinuity is introduced using dedicated basis functions for both $T$ \cite{Schnaubelt_2023aa} and $\vec{H}$ \cite{Schnaubelt_2023ab}, rather than on the mesh level as proposed in \cite{Alves_2021aa}. Thanks to this choice, the thermal and magnetodynamic TSA can use the same mesh avoiding the need for interpolation between different meshes.

As shown in Fig.~\ref{fig:virtDisc}, $T_+$ and $T_-$ with support restricted to one of the two sides of $\Gamma_\text{c,cl}$ are introduced as well as corresponding test functions $T'_+$ and $T'_-$. The procedure is repeated with the magnetic field strength yielding $\vec{H}_+$, $\vec{H}_-$, $\vec{H}'_+$, $\vec{H}'_-$. This approach leads to additional surface contributions in the weak formulations \eqref{eq:thermal} and \eqref{eq:mqs} \cite{Schnaubelt_2023aa, Schnaubelt_2023ab}. These contributions are used to consider an interface condition on $\Gamma_\text{c,cl}$. 

This interface condition is built by an \emph{internal FE discretization} of the weak formulations \eqref{eq:thermal} and \eqref{eq:mqs}. This discretization takes place in the internal domain $\hat{\Omega}_\text{c,cl}$ which is a representation of the T2TCL $\Omega_\text{c,cl}$ (see Fig.~\ref{fig:virtDisc}). To build the internal FE discretization, a local coordinate system $\left(\vec{u}, \vec{v}, \vec{w}\right)$ is used with the same orientation as for the anisotropic material properties, i.e., with $\vec{u}$ in tangential direction along the spiral, $\vec{v}$ in axial direction and $\vec{w}$ along the normal direction of $\Gamma_\text{c,cl}$. The domain $\hat{\Omega}_\text{c,cl}$ is subdivided into $N$ layers $\hat{\Omega}_\text{c,cl}^{(k)}$ for $k= 1, ..., N$ with $\hat{\Omega}_\text{c,cl} = \bigcup_{k = 1}^N \hat{\Omega}_\text{c,cl}^{(k)}$ and $\hat{\Omega}_\text{c,cl}^{(k)} := \Gamma_\text{c,cl}^{(k)} \times \left[w_{k-1}, w_{k} \right]$. In each $\hat{\Omega}_\text{c,cl}^{(k)}$, we make a product ansatz 
\begin{align}
    \vec{H}|_{\hat{\Omega}_\text{c,cl}^{(k)}} (u, v, w, t) &= \sum_{j = k-1}^k \vec{H}_{\text{t}}^j (u, v, t) \, \chi_j(w) \label{eq:ansatzH}\\ 
    T|_{\hat{\Omega}_\text{c,cl}^{(k)}} (u, v, w, t) &= \sum_{j = k-1}^k T^j (u, v, t) \, \chi_j(w),  \label{eq:ansatzT}
\end{align}
where $\vec{H}_{\text{t}}^j$ is the tangential magnetic field strength on $\Gamma_\text{c,cl}^{(j)}$ and $\chi_j(w)$ the basis function along $\vec{w}$. The latter can be chosen to match the simulation needs (including higher order functions). In this work, we use first-order Lagrange basis functions
\begin{align*}
     \chi_{k-1}(w) = \frac{w_{k} - w}{w_k - w_{k-1}} \quad \text{and} \quad \chi_k(w) = \frac{w - w_{k - 1}}{w_k - w_{k-1}},
\end{align*}
for both problems. Let us note that the functions do not have to be the same for the thermal and magnetodynamic problem. However, by choosing them to be the same, one simplifies the implementation of the 1D FE integrals of the TSA formulation since it allows to reuse existing routines for their integration. The test functions $T'$ and $\vec{H}'$ are discretized with the same ansatz. This leads to a decomposition of the internal problem into surface integrals
on $\Gamma_\text{c,cl}^{(k)}$ and 1D FE integrals on $\left[w_{k-1}, w_{k} \right]$.
The final expressions for the terms in weak formulation \eqref{eq:thermal} and \eqref{eq:mqs} are found in \cite{Schnaubelt_2023aa, Alves2022, Schnaubelt_2023ab} except for the Joule losses term inside $\hat{\Omega}_\text{c,cl}$, that is,
\begin{equation}
   \left( \rho \, \vec{J} \cdot \vec{J}, T' \right)_{\hat{\Omega}_\text{c,cl}} = \sum_{k = 1}^N \left( \rho \, \vec{J} \cdot \vec{J}, T' \right)_{\hat{\Omega}_\text{c,cl}^{(k)}}.
\end{equation}
Its derivation is detailed in Appendix~\ref{appendix:JouleLosses}, which constitutes the first part of the coupling between magnetodynamics and heat equation. To appropriately account for the second part, i.e., the temperature- and field-dependence of the material parameters, the ansatzes \eqref{eq:ansatzH} and \eqref{eq:ansatzT} need to be used in the 1D FE integrals as derived in the appendix, alongside an appropriate numerical quadrature. In this work, Gaussian integration is used as it is readily available in the FE software. To account for the increased volume of $\Omega_\text{c,b,TSA}$ compared to $\Omega_\text{c,b}$, a material scaling is introduced as detailed in Appendix~\ref{app:modifiedMaterial}.

\section{Implementation and Simulation Setup}
\label{sec:setup}

The magneto-thermal model, with and without the TSA, is implemented in the open-source framework GetDP 3.5 \cite{Dular_1998aa} using the Gmsh 4.11.0 \cite{Geuzaine_2009ab} application programming interface to create the geometry and mesh. It also creates the cohmology basis functions required for the $\vec{H}-\phi$ formulation \cite{pellikka2013}. Lowest order basis functions are used and as the T2TCL consists of only one material, the TSA is used with $N = 1$. An adaptive implicit Euler scheme is used for time integration \cite{Ascher_1998aa}. 

\begin{figure}[!tbh]
    \centering
    \scalebox{1.0}{\input{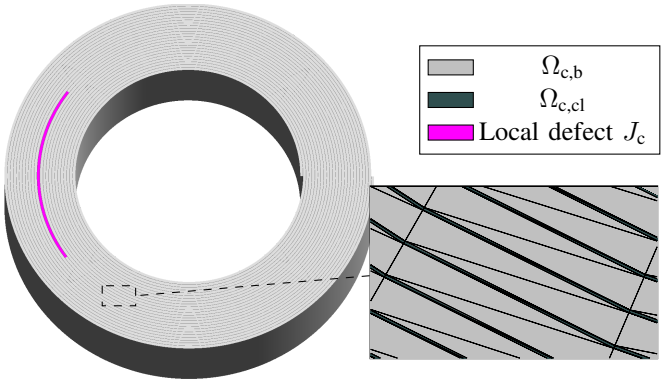}}
    \caption{Local defect and zoom on the mesh of the volumetric T2TCL.}
    \label{fig:localDefect}
\end{figure}

All important parameters regarding the coil geometry and simulation setup are given in Table~\ref{tab:coilparams}. A local defect with $J_\text{c} = 0$ is introduced to highlight the current diversion across the T2TCL, a mechanism that increases the thermal stability of NI coils. The local defect is depicted in Fig.~\ref{fig:localDefect} which also shows the mesh of the geometry with volumetric T2TCL.

\begin{table}[!tbh]
    \centering
        \caption{Summary of model parameters.}
    \label{tab:coilparams}
    \begin{tabular}{cc}
    \hline \hline
        Description & Value \\ \hline
         {Power law $n$-value} & {30} \\ 
        {Critical current $I_\text{c}$} &   $I_\text{c} \left(\norm{\vec{B}}, T, \theta \right)$ \cite{steam_material_library}\\ 
        {Critical current at $\vec{B} = 0$ and $T = \SI{15}{\kelvin}$} & {\SI{780}{\ampere}}\\ 
        {ReBCO thickness} & {\SI{1.5}{\micro \meter}} \\ 
        \{Cu, Hastelloy\textsuperscript{\textregistered}, Ag\} thickness  &  \{42, 75, 1.5\} \si{\micro \meter} \cite{steam_material_library} \\
        Source current $I_\text{src}$ & Powering cycle (Fig.~\ref{fig:centralField}) \\
        Number of turns $N_\text{t}$ & 24 \\
        Inner radius & \SI{5}{\milli \meter} \\ 
        Bare conductor width $w_\text{t}$ & {\SI{4}{\milli \meter}} \\ 
        T2TCL thickness $\text{th}_\text{cl}$ & \SI{10}{\micro \meter} \\
        {T2TCL resistivity}  $\rho_\text{cl}$ & {\SI{1.12E-4}{\ohm \meter} \cite{9369013}}\\
        {T2TCL thermal conductivity}  $\kappa_\text{cl}$ & Stainless steel \cite{steam_material_library} \\
        {T2TCL heat capacity}  $C_\text{V, cl}$ & Stainless steel \cite{steam_material_library} \\
        Initial temperature $T$ at $t = \SI{0}{\second}$ & \SI{15}{\kelvin}\\
        Local defect with $J_\text{c} = 0$ & Turn 12.4 to 12.6 (Fig.~\ref{fig:localDefect})\\
        Current lead material & Cu \cite{steam_material_library} with $\text{RRR} = 100$\\
        Temperature boundary condition $g$ &  $ g = \SI{15}{\kelvin} \quad \forall t$\\
        \hline
    \end{tabular}

\end{table}

\section{Powering Cycle Simulation With Local Defect}
\label{sec:numExpCoupled}

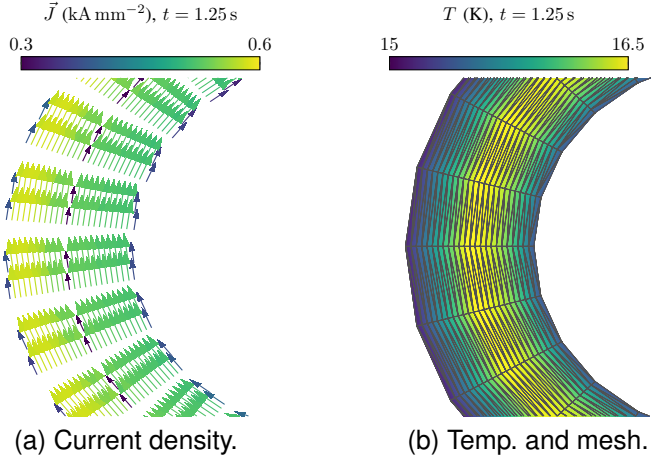
\begin{figure}[htb]
     \centering
        \subfloat{
            \centering
            \scalebox{0.7}{\begin{tikzpicture}
    \begin{axis}[
    	 hide axis,
        scale only axis,
        height=0pt,
        width=.25\textwidth,
    	title style={align=center}, 
    	title={$\vec{J}$ (\si{\kilo \ampere \per \milli \meter \squared}), $t = \SI{1.25}{\second}$},
    	colorbar,
    	colormap/viridis,
    	colorbar horizontal,
    	colorbar style={
    	    point meta min=0.3, point meta max=0.6,
    		width=.25\textwidth,
    		height=0.25cm, 
    		xticklabel pos=upper, 
    		xtick={0.3, 0.6}
    	}, 
    	    xtick={0.3, 0.6}
    	]
    	  \addplot[draw=none]coordinates{(0,1)};
    \end{axis}
\end{tikzpicture}%
}
       } \hfill 
        \subfloat{
            \centering
            \scalebox{0.7}{\begin{tikzpicture}
    \begin{axis}[
    	 hide axis,
        scale only axis,
        height=0pt,
        width=.25\textwidth,
    	title style={align=center}, 
    	title={$T$ (K), $t = \SI{1.25}{\second}$},
    	colorbar,
    	colormap/viridis,
    	colorbar horizontal,
    	colorbar style={
    	    point meta min=15, point meta max=16.5,
    		width=.25\textwidth,
    		height=0.25cm, 
    		xticklabel pos=upper, 
    		xtick={15, 16.5}
    	}, 
    	    xtick={15, 16.5}
    	]
    	  \addplot[draw=none]coordinates{(0,1)};
    \end{axis}
\end{tikzpicture}%
}
       }\\[-0.25cm]
       \setcounter{subfigure}{0}
       \subfloat[Current density.]{
            \centering
            \includegraphics[width=.18\textwidth, height=4.5cm]{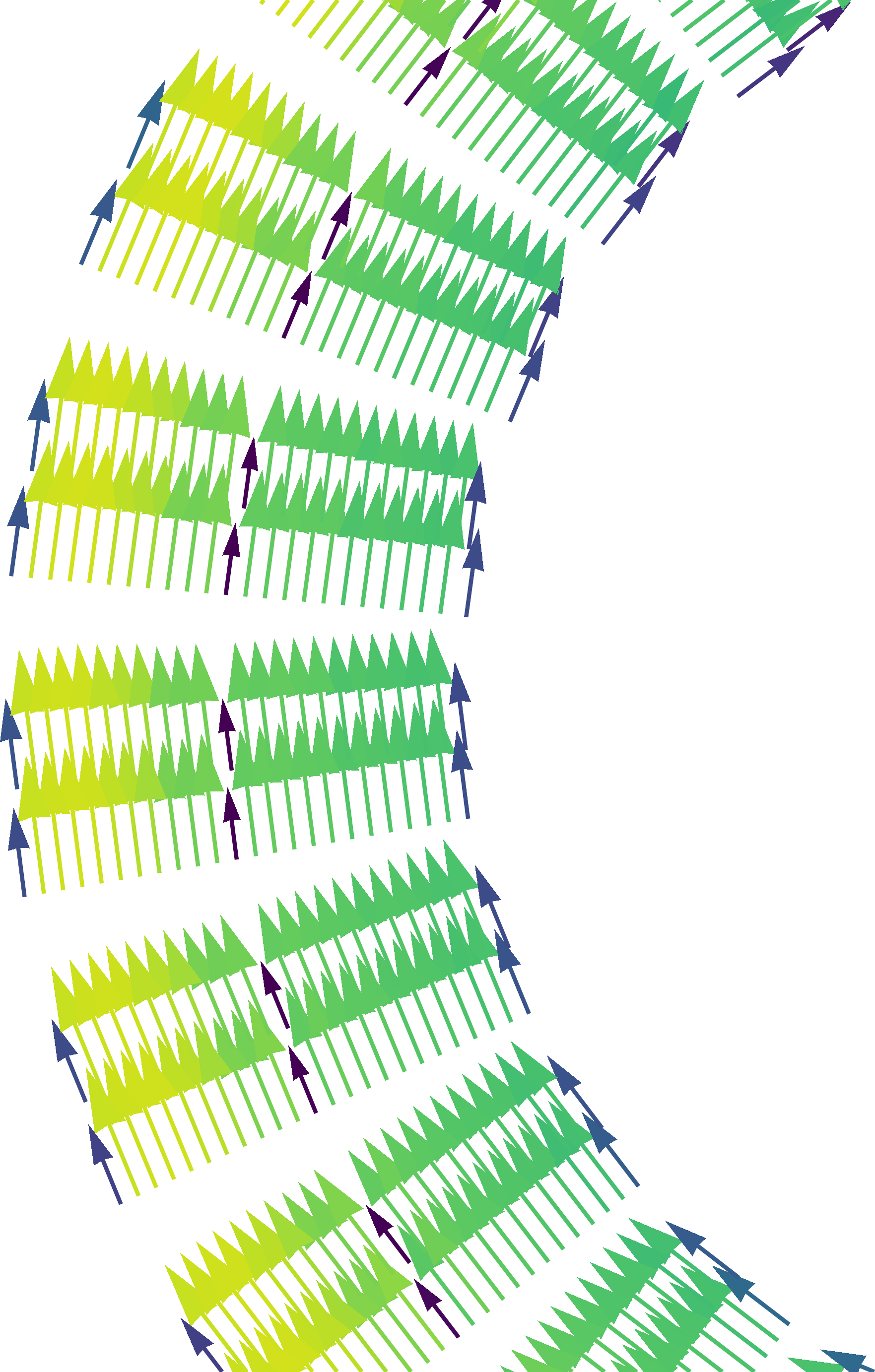}
            \label{fig:currentDiversion}
       } \hfill
        \subfloat[Temp. and mesh.]{
            \centering
            \includegraphics[width=.18\textwidth, height=4.5cm]{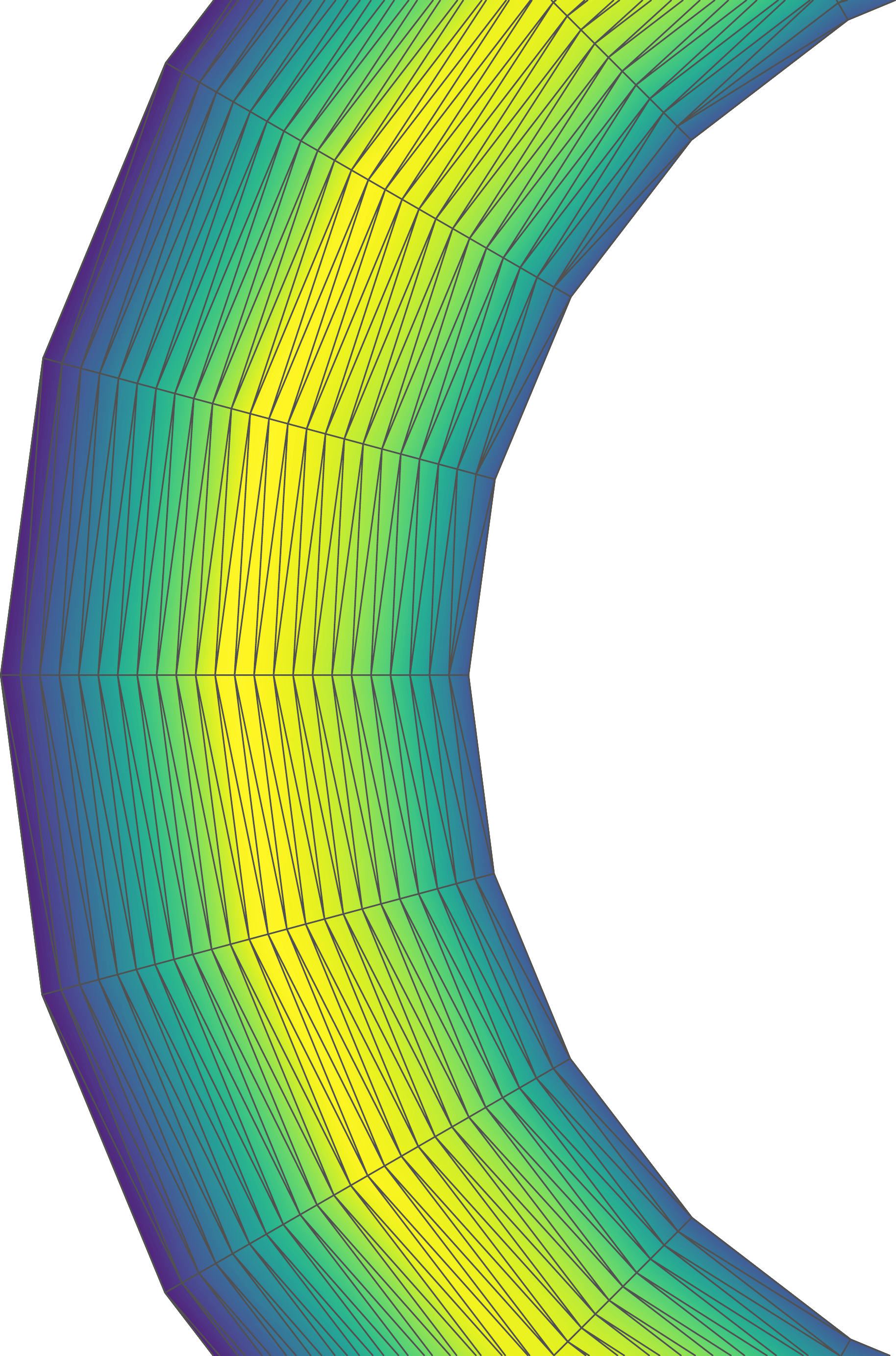}
            \label{fig:temp}
       }
        \caption{Current by-pass and temperature distribution at $t = \SI{1.25}{\second}$. 
        }
\end{figure}

For verification, the powering cycle of the NI  coil is simulated using three models; \emph{i) vol}: a fine-meshed model with volumetric T2TCL, \emph{ii) TSAf}: a T2TCL TSA model which uses the same mesh structure inside $\Omega_\text{c,b}$ as the \emph{vol} model and \emph{iii) TSAc}: a coarse-meshed T2TCL model (see Fig.~\ref{fig:currentDiversion} for the TSAc mesh). It has been proven difficult to use coarsely meshed models with volumetric T2TCL, either because of failure to automatically create suitable meshes or convergence issues due to the high aspect ratio of mesh elements. The TSA T2TCL models have been more robust and the aforementioned problems did not occur. Thus, it is possible to use coarse mesh with the TSA while achieving an accurate solution.

\begin{figure}[htb]
         \centering
          \begin{tikzpicture}
 
	\begin{axis}[cycle list={[
	colormap/viridis, 
	colors of colormap={0, 0, 400, 800}]}, width=.425\textwidth,height=3.75cm,
	xlabel={Time (s)},
	ylabel={Current $I_\text{src}$ (A)}, 
	ylabel near ticks, 
	axis y line*=left,
    xtick={0, 0.5, 1, 1.5, 2, 2.5}, 
    ytick={0, 175, 350}, 
    yticklabels={0, 175, 350},
	  xtick scale label code/.code={}
		 ]
       \addplot+[mark=none, densely dashed,black] table[x index={0}, y index={1}, col sep=comma, skip first n = 1]{\currentRef}; \label{input_current_centralField}
        


	\end{axis}

	\begin{axis}[axis y line*=right,
		cycle list={[
	colormap/viridis, 
	colors of colormap={0, 300, 600}]}, width=.425\textwidth,height=3.75cm,
	ylabel={Flux dens. $B_\text{z, c}$ (T)}, 
	ylabel near ticks, 
	legend columns=1, 
	legend style={font=\footnotesize, at={(0.99, 0.99)}, anchor=north east}, 
	label style={font=\small}, 
    legend image post style={mark indices={}},
    ytick={0, 0.3, 0.6}, 
    legend cell align={left}, 
    xticklabels={},
     	  xtick scale label code/.code={}]
		 \addlegendimage{/pgfplots/refstyle=input_current_centralField}

          \addplot+[mark=none
           , y filter/.code={\pgfmathparse{\pgfmathresult*(-1)}\pgfmathresult}
          ] table[x index = {1}, y index = {5}]{\bFieldRef};
          
          \addplot+[mymark={triangle}{solid, opacity=1}, opacity=0
          ] table[x index = {1}, y index = {5}]{\bFieldTSA};

          \addplot+[mymark={square}{solid, opacity=1}, opacity=0
         , y filter/.code={\pgfmathparse{\pgfmathresult*(-1)}\pgfmathresult}
          ] table[x index = {1}, y index = {5}]{\bFieldVol};

	      \legend{$I_\text{src}$, $B_\text{c,vol} $, $B_\text{z,c,TSAf}$, 
       $B_\text{c,TSAc}$}
	\end{axis}
\end{tikzpicture}%
         \caption{{The source current $I_\text{src}$ and the central axial magn. flux density $B_{z, \text{c}}$.}}
         \label{fig:centralField}
\end{figure}
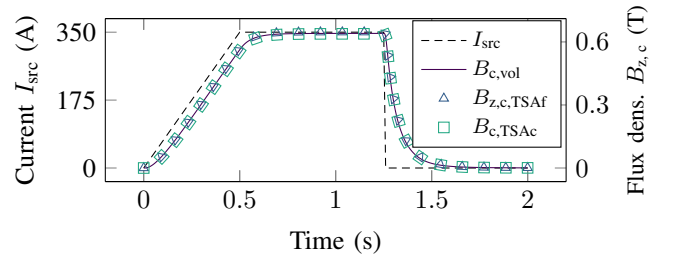

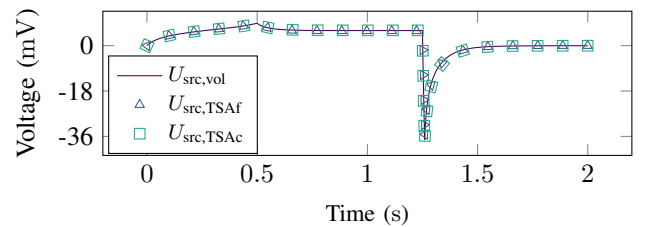
\begin{figure}[htb]
         \centering
         \begin{tikzpicture}
    \begin{axis}[cycle list={[
    	colormap/viridis, 
    	colors of colormap={0, 300, 600}]}, 
        width=.475\textwidth, height=3.5cm,
    	ylabel={Voltage (\si{\milli \volt})},
        xlabel={Time (\si{\second})}, 
    	legend columns=1, 
    	legend style={font=\footnotesize, at={(0.01, 0.01)}, anchor=south west}, 
    	label style={font=\small}, 
        legend image post style={mark indices={}}, 
        legend cell align={left}, 
     	  ytick={0, 0.018, 0.036},
     	  yticklabels={0, -18, -36},
         scaled y ticks={real:1E-3},
         ytick scale label code/.code={}, 
         y post scale=-1, 
             ylabel near ticks
          ]
    
              \addplot+[mark=none, y filter/.code={\pgfmathparse{\pgfmathresult*(-1)}\pgfmathresult}] table[x index = {0}, y index = {1}, skip first n=1 
              ]{\voltageRef};
              
              \addplot+[mymark={triangle}{solid, opacity=1}, opacity=0, y filter/.code={\pgfmathparse{\pgfmathresult*(-1)}\pgfmathresult}] table[x index = {0}, y index = {1}, skip first n=1]{\voltageTSA};
               
              \addplot+[mymark={square}{solid, opacity=1}, opacity=0,  y filter/.code={\pgfmathparse{\pgfmathresult*(-1)}\pgfmathresult}] table[x index = {0}, y index = {1}, skip first n=1]{\voltageVol};

                        
    	      \legend{$U_\text{src,vol}$, $U_\text{src,TSAf}$, $U_\text{src,TSAc}$}
    \end{axis}
\end{tikzpicture}%
         \caption{{ Coil terminals voltages $U_\text{src}$ for all models.}}
         \label{fig:voltages}
\end{figure}

\begin{figure}[htb]
         \centering
         \begin{tikzpicture}
    \begin{axis}[
        cycle list={[
    colormap/viridis, 
    colors of colormap={0, 300, 600}]}, width=.475\textwidth,height=3.5cm,
    ylabel={Temp. $T_\text{d}$ (K)}, 
    ylabel near ticks, 
    xlabel={Time (\si{\second})}, 
    legend columns=1, 
    legend style={font=\footnotesize, at={(0.99, 0.99)}, anchor=north east}, 
    label style={font=\small}, 
    legend image post style={mark indices={}},
    legend cell align={left}, 
          xtick scale label code/.code={}]
    
          \addplot+[] table[x index = {1}, y index = {5}]{\temperatureRef};
    
          \addplot+[mymark={triangle}{solid, opacity=1}, opacity=0] table[x index = {1}, y index = {5}]{\temperatureTSA};
          
          \addplot+[mymark={square}{solid, opacity=1}, opacity=0] table[x index = {1}, y index = {5}]{\temperatureVol};
    
          \legend{$T_\text{d,vol}$, $T_\text{d,TSAf}$, $T_\text{d,TSAc}$}
    \end{axis}
\end{tikzpicture}%
         \caption{{ Temperature at the center of the local defect for all models.}}
         \label{fig:temperature}
\end{figure}
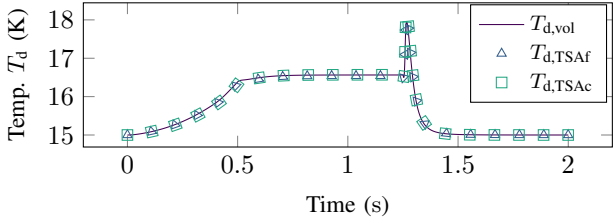

The applied source current and axial central magnetic flux density are depicted in Fig.~\ref{fig:centralField} for all three models. The magnetic field is delayed w.r.t. the source current due to the radial currents and decays exponentially after the sudden discharge. All three models are in good agreement. The same is true for voltage between the coil terminals which is shown in Fig.~\ref{fig:voltages}. During charging and discharging, the absolute value of the voltage is highest as the radial currents flow in the highly resistive T2TCL. It does not decay to zero during the current plateau as \emph{i)} the current leads are made of copper and \emph{ii)} radial currents are crossing the T2TCL to divert around the local defect. The current diversion is depicted in Fig.~\ref{fig:currentDiversion} and the local temperature hot spot at the defect position in Fig.~\ref{fig:temp}.
Thanks to the current diversion and the cooling applied at the coil terminals using boundary condition \eqref{eq:dir}, the local defect does not cause a thermal runaway as seen in Fig.~\ref{fig:temperature}. During charging, the temperature increases due to the currents across the T2TCL. During the current plateau, the heat generated by the currents across the T2TCL is removed by the cooling yielding a constant temperature. During the fast discharge, the large T2TCL currents first cause a significant temperature increase which is then cooled down subsequently.

Having shown the accuracy of the TSA approach, the number of DoF and solution time are shown in Table~\ref{tab:num_dof}. Let us note that the implementation of the finite difference derivative makes heavy use of GetDP's scripting language. While this is convenient for the implementation, this increases significantly the time required for assembly of the linear systems and thus also the total solution time required. It does not have an influence on the relative comparison of solution times between the models but only on the absolute timing. The TSAc model has 3.5 times fewer DoF and 3.4 times shorter solution time than the vol model. For vanishing T2TCL thickness (i.e., directly touching turns), it becomes increasingly difficult to use a volumetric T2TCL \cite{Schnaubelt_2023aa}. The TSA then is the only practical method that is expected to lead to an accurate solution. Furthermore, it simplifies the meshing. 

\begin{table}[H]
    \centering
        \caption{{ Number of DoF and total solution time of the three models.}}
    \label{tab:num_dof}
    \begin{tabular}{cccc}
    \hline \hline
    Description & \emph{vol} & \emph{TSAf} & \emph{TSAc} \\ \hline 
    Number of DoF thermal system &  14907 & 14896 & 5229 \\
    Number of DoF magnetodynamic system &  49740 & 39678 & 13090 \\
    Total solution time in h & 17.4 & 16.9 & 5.05 \\
    \hline
    \end{tabular}
\end{table}
\section{Conclusion}
\label{sec:conclusion}



A coupled magneto-thermal TSA has been presented to model the T2TCL in the FE analysis of HTS NI coils in 3D. The formulation has been implemented in the open-source FE framework Gmsh/GetDP and a reference implementation has been made publicly available. The numerical results of a powering cycle of a model NI coil have been verified against a model with volumetric T2TCL showing excellent agreement. The TSA leads to a significantly reduced number of DoF, solution time and simplifies the creation of high-quality meshes. Its robustness and efficiency enable automated models resolving each turn to capture local effects like quench. 

\appendices
\section{Derivation of Joule Loss Expression}
\label{appendix:JouleLosses}%
From the representation of $\vec{H}$ in $\hat{\Omega}_\text{c,cl}^{(k)}$ in \eqref{eq:ansatzH},
we find that
\begin{align}
\vec{J}|_{\hat{\Omega}_\text{c,cl}^{(k)}} &= \curl \vec{H}|_{\hat{\Omega}_\text{c,cl}^{(k)}}
    = \curl \left[\sum_{j = k-1}^k \vec{H}_{\text{t}}^j \, \chi_j \right], \\
    &=  \sum_{j = k-1}^k 
    \gradient \chi_j \times \vec{H}_{\text{t}}^j + \chi_j \curl
    \vec{H}_{\text{t}}^j, \\
    &=  
    \vec{w} \times \frac{\vec{H}_{\text{t}}^{k} - \vec{H}_{\text{t}}^{k - 1}}{w_k - w_{k-1}} + \sum_{j = k-1}^k  \chi_j \curl
    \vec{H}_{\text{t}}^j. \label{eq:joul}
\end{align}
Note that first summand is orthogonal to the $\vec{w}$-direction while the second is in $\vec{w}$-direction. Using \eqref{eq:joul}, the ansatz \eqref{eq:ansatzT} for $T'$ and $l \in \{k-1,k\}$, the Joule losses in the TSA evaluate to 
\begin{align*}
    &\left( \rho \, \vec{J} \cdot \vec{J}, T' \right)_{\hat{\Omega}_\text{c,cl}^{(k)}}  \\ 
    = & \left( \rho   \norm{\frac{\vec{H}_{\text{t}}^{k} - \vec{H}_{\text{t}}^{k - 1}}{w_k - w_{k-1}}} ^2  , T_l' \chi_l \right)_{\hat{\Omega}_\text{c,cl}^{(k)}} + \\
    &\left( \rho    \sum_{i = k-1}^k  \left( \chi_i \curl
    \vec{H}_{\text{t}}^i \right) \cdot  \sum_{j = k-1}^k  \left(  \chi_j \curl
    \vec{H}_{\text{t}}^j \right) ,  T_l' \chi_l  \right)_{\hat{\Omega}_\text{c,cl}^{(k)}} \\
    =  & \left< f_l  \norm{\frac{\vec{H}_{\text{t}}^{k} - \vec{H}_{\text{t}}^{k - 1}}{w_k - w_{k-1}}} ^2  , T_l'  \right>_{\Gamma_\text{c,cl}^{(k)}} + \\
    &  \sum_{i = k-1}^k \sum_{j = k-1}^k
    \left< c_{ijl}
          \curl \vec{H}_{\text{t}}^i
    \cdot \curl
      \vec{H}_{\text{t}}^j,  T_l'   \right>_{\Gamma_\text{c,cl}^{(k)}}.
\end{align*}
Herein,  $\left<{\cdot},{\cdot}\right>_{\Gamma}$  denotes the surface integral of the scalar product of its two arguments on $\Gamma$. We introduced the 1D FE integrals
\begin{equation*}
    f_l =  \int_{w_{k-1}}^{w_k} \rho \chi_l \, \text{d}w \quad \text{and}  \quad c_{ijl} =
    \left(  \int_{w_{k-1}}^{w_k} \rho \chi_i \chi_j \chi_l \, \text{d}w \right).
\end{equation*}
This concludes the decomposition into surface integrals on $\Gamma_\text{c,cl}^{(k)}$ and 1D FE integrals in $\left[ w_{k-1}, w_k\right]$, alleviating the need for a volumetric mesh representation of $\Omega_\text{c,cl}$. 

\section{Material Scaling for TSA Model}
\label{app:modifiedMaterial}

Using the scaling factor 
\begin{equation}
    p = \frac{|\Omega_{\text{c,cl}}| + |\Omega_{\text{c,b}}|}{|\Omega_{\text{c,b}}|} = 1 + \frac{|\Omega_{\text{c,cl}}|}{|\Omega_{\text{c,b}}|} = \frac{|\Omega_{\text{c,b,TSA}}|}{|\Omega_{\text{c,b}}|}, 
\end{equation}
the material parameters of $\Omega_\text{c,b,TSA}$ are scaled by
\begin{alignat*}{3}
    \kappa_{uu}|_{\Omega_\text{c,b,TSA}} &= p^{-1} \, \kappa_{uu}|_{\Omega_\text{c,b}}, \qquad 
    &\kappa_{ww}|_{\Omega_\text{c,b,TSA}} &= p \, \kappa_{ww}|_{\Omega_\text{c,b}},\\
    \rho_{uu}|_{\Omega_\text{c,b,TSA}} &= p \, \rho_{uu}|_{\Omega_\text{c,b}}, \qquad
    &\rho_{ww}|_{\Omega_\text{c,b,TSA}} &= p^{-1} \, \rho_{ww}|_{\Omega_\text{c,b}},\\  
    C_\text{V}|_{\Omega_\text{c,b,TSA}} &= p^{-1} \,  C_\text{V}|_{\Omega_\text{c,b}}, \qquad &J_\text{c}|_{\Omega_\text{c,b,TSA}} &= p^{-1} \, J_\text{c}|_{\Omega_\text{c,b}}.
\end{alignat*}

\printbibliography

\end{document}